# Strain-driven elastic and orbital-ordering effects on thickness-dependent properties of manganite thin films

I. C. Infante*, F. Sánchez, and J. Fontcuberta
*Institut de Ciència de Materials de Barcelona-CSIC, Campus UAB, 08193 Bellaterra, Catalonia, Spain*

M. Wojcik and E. Jedryka
*Institute of Physics, Polish Academy of Sciences, Al. Lotnikow 32/46, 02 668 Warszawa, Poland*

S. Estradé and F. Peiró
*EME/CeRMAE/IN2UB, Dept. d'Electrònica, Universitat de Barcelona, 08028 Barcelona, Catalonia, Spain*

J. Arbiol
*EME/CeRMAE/IN2UB, Dept. d'Electrònica and TEM-MAT, Serveis Cientificotècnics, Universitat de Barcelona, 08028 Barcelona, Catalonia, Spain*

V. Laukhin
*Institut de Ciència de Materials de Barcelona-CSIC, Campus UAB, 08193 Bellaterra, CAT, Spain and Institut Català d'Investigació i Estudis Avançats (ICREA), Barcelona, Catalonia, Spain*

J. P. Espinós
*Instituto de Ciencia de Materiales de Sevilla ICMSE (CSIC-USE), c/ Américo Vespucio s/n, 41092 Sevilla (Spain)*

We report on the structural and magnetic characterization of (110) and (001) $La_{2/3}Ca_{1/3}MnO_3$ (LCMO) epitaxial thin films simultaneously grown on (110) and (001)$SrTiO_3$ substrates, with thicknesses $t$ varying between 8 nm and 150 nm. It is found that while the in-plane interplanar distances of the (001) films are strongly clamped to those of the substrate and the films remain strained up to well above $t \approx 100$ nm, the (110) films relax much earlier. Accurate determination of the in-plane and out-of-plane interplanar distances has allowed concluding that in all cases the unit cell volume of the manganite reduces gradually when increasing thickness, approaching the bulk value. It is observed that the magnetic properties (Curie temperature and saturation magnetization) of the (110) films are significantly improved compared to those of (001) films. These observations, combined with $^{55}$Mn-nuclear magnetic resonance data and X-ray photoemission spectroscopy, signal that the depression of the magnetic properties of the more strained (001)LCMO films is not caused by an elastic deformation of the perovskite lattice but rather due to the electronic and chemical phase separation caused by the substrate-induced strain. On the contrary, the thickness dependence of the magnetic properties of the less strained (110)LCMO films are simply described by the elastic deformation of the manganite lattice. We will argue that the different behavior of (001) and (110)LCMO films is a consequence of the dissimilar electronic structure of these interfaces.

# I. INTRODUCTION

The manganese perovskites $R_{1-x}A_xMnO_3$, where R is a trivalent rare earth and A is a divalent alkaline earth, have been the focus of a colossal research activity due to their half-metallic ferromagnetic character and the observation of a huge magnetoresistance. The exploitation of the full spin polarization in advanced magnetic applications, such as magnetic tunnel junctions (MTJ), involves the fabrication of epitaxial heterostructures, where a $R_{1-x}A_xMnO_3$ electrode is grown on top of single crystalline substrates. It is thus important to understand and to tailor the effects induced by the substrate on the magnetic film. One of the most well documented substrate-induced effects on the properties of manganite thin films is the strong dependence of the Curie temperature ($T_C$) and saturation magnetization ($M_S$) on film thickness (*t*).

It is commonly found that $T_C$ and $M_S$ in thin films are depressed compared to the bulk counterparts [1-5] and although these effects were proposed to be the consequence of the epitaxial strain acting on the lattice [6], conclusive response remains elusive [7,8]. Millis *et al.* [6] suggested that substrate induced strain has two distinguishable consequences: (i) in analogy to what has been experimentally observed in bulk materials [9,10], a uniform compression of the manganite unit cell would lead to a broadening of the conduction band, a reinforcement of the $Mn^{3+}$-$Mn^{4+}$ electron transfer and a weakening of the electron-phonon coupling, thus leading to an increase of $T_C$, and (ii) a biaxial distortion would strengthen the Jahn-Teller distortion and would produce a subsequent increase of the splitting of the $e_g$ electron bands, eventually reinforcing the tendency to charge localization. Within this scenario, the dependence of $T_C$ on the substrate-induced strain can be written as:

$$T_C(\varepsilon) = T_C(\varepsilon = 0)(1 - \alpha \varepsilon_B - \frac{1}{2}\beta \varepsilon_{bi}^2), \tag{1}$$

where $\varepsilon_B = (1/3)(\varepsilon_{11} + \varepsilon_{22} + \varepsilon_{33})$ is the bulk strain and $\varepsilon_{bi}$ is the biaxial strain, which has two components: $\varepsilon^{in}_{bi} = (1/2)(\varepsilon_{11} - \varepsilon_{22})$, and $\varepsilon^{out}_{bi} = (1/4)(2\varepsilon_{33} - \varepsilon_{11} - \varepsilon_{22})$. The indexes (11, 22) refer to two orthogonal in-plane directions and (33) refers to the out-of-plane direction. $\alpha = (1/T_C) dT_C/d\varepsilon_B$ is related to the unit-cell volume variations under strain ($\Delta V/V = 3 \varepsilon_B$) and $\beta$ accounts for the Jahn-Teller energy splitting. It follows from eq. (1) that whereas the biaxial distortion always leads to a reduction of $T_C$, this is not the case for the bulk distortion, which can be either positive or negative depending on the sign of $\Delta V/V$ under strain. Experimentally it has been found that in almost all the cases the $T_C$ of films grown on single crystalline substrates is lower than its corresponding bulk value, irrespectively to the sign of the bulk strain, thus suggesting the major relevance of the biaxial component of the strain on the $T_C$ values.

Assuming an elastic unit-cell deformation, the out-of-plane strain is related to the in-plane strain by the elastic constants of the lattice, and differs depending on the particular orientation of the substrate. For a cubic substrate and a (001) plane, $\varepsilon_{33} = -\nu / (1-\nu) (\varepsilon_{11} + \varepsilon_{22})$ where $\nu$ is the Poisson ratio given by $\nu = c_{11} / (c_{11} + c_{12})$ being $c_{11}$ and $c_{12}$ the appropriate elastic constants [11]. When $\nu = 1/2$, the unit cell volume is preserved under strain, otherwise it changes.

Biaxial strain is most commonly evaluated [2,4,8] from the measured out-of-plane cell parameters of the manganite film assuming an *ad-hoc* value for the Poisson ratio $\nu$. However, Maurice *et al.* [5] measured the in-plane and out-of-plane unit cell parameters of $La_{2/3}Sr_{1/3}MnO_3$ epitaxial films on (001)$SrTiO_3$ (STO) substrates and derived $\nu = 0.34$. This implies that the unit cell volume of these strained films increases compared to the bulk value. Rao *et al.* [12] in $La_{0.8}Ca_{0.2}MnO_3$ epitaxial films and Bibes *et al.* in $La_{2/3}Ca_{1/3}MnO_3$ (LCMO) ones [13] found a similar trend when films were grown on (001)STO substrates. It is a remarkable observation that whereas the in-plane parameters are clamped to those of the substrate up to large film thickness, the out-of-plane parameter relaxes gradually when increasing thickness [5,12,13], indicating that the Poisson ratio changes with thickness. Thus, a simple model of elastic deformation under biaxial strain of the manganite lattice does not hold to this system.

The common assumption of elastic deformation of lattices under strain was proven to be true in simple systems such as heteroepitaxies of metals and some semiconductors, although we recall that in some cases (SiGe, for instance) it has been recently shown that strain is relaxed via compositional changes [14]. It remains to be demonstrated that the assumptions of elastic unit cell deformation, constant unit cell volume and invariable composition as a function of film thickness still hold in the case of manganese perovskites. It is important to note that the elastic energy induced by epitaxial strain can also be relaxed in other ways depending on the real structure of the bulk manganite (rhombohedral or orthorhombic), the growth mode or the eventual chemical fluctuations [5].

Moreover, it has been proposed that interface effects, either due to symmetry breaking or strain, are prone to induce chemical [15] or electronic phase separation [13]. To what extent these interface effects are at the origin or contribute to the fast fall-off with temperature of the magnetoresistance in MTJ's [16] remains undisclosed. The electronic phase separation could also be driven by electronic interaction with substrates. As it has been recently pointed out [17], charge density gradients or polarity discontinuities across interfaces in oxide-based heterostructures can be a driving force for charge redistribution in the oxide layers. On the other hand, the stabilization of A or C-type orbital order in strained manganite layers has been observed by means of linear and magnetic circular dichroism in X-ray absorption spectroscopy [18-20]. Besides, Brey [21] has recently shown that the interface between half-metallic manganites and insulators tends to be charge-depleted, favoring the weakening of the ferromagnetic interaction and the stabilization of orbital-ordered insulating states.

The purpose of this study is to compare the structure and magnetic properties of La$_{2/3}$Ca$_{1/3}$MnO$_3$ (LCMO) thin films of different thickness grown on (001) and (110) single crystalline STO substrates. Pseudo-cubic lattice parameters of LCMO and STO are a$_{LCMO}$ = 0.3863 nm and a$_{STO}$ = 0.3905 nm, respectively, and there is the same lattice mismatch (tensile strain) of f = (a$_{STO}$-a$_{LCMO}$)/a$_{LCMO}$ = 1.1 %. However, (001) and (110) STO free surfaces differ in several fundamental aspects. For instance: (110) surfaces of STO and of R$_{1-x}$A$_x$MnO$_3$ do not display square symmetry but rectangular; the atomic sequences along the two in-plane orthogonal directions ([001] and [1-10]), of (110) surfaces radically differ: Mn-O-Mn are separated by a distance $a_p$ along the [001] direction ($a_p$ being the pseudo-cubic unit cell length), and Mn-Mn are separated by $a_p\sqrt{2}$ along the [1-10] direction (Fig. 1). As a consequence, one could expect that the elastic properties of the manganite layer would differ along these two directions, leading to the possibility of anisotropic strain relaxation mechanisms which will open the possibility to tailor the in-plane unit cell parameters. Thus, although the film-substrate lattice mismatch is identical for (001) and (110) textured films, the resulting strains can be radically different.

Manganite films and heterostructures with orientations other than (001) have been scarcely investigated. Nevertheless, it has been shown that (110) oriented manganite films present a uniaxial in-plane magnetic anisotropy [22-24], the anisotropic magnetoresistance being dependent on the in-plane applied current direction [24]. $^{55}$Mn nuclear magnetic resonance (NMR) experiments on (110) oriented LCMO films on (110)STO have shown the resonance line of mixed valent Mn$^{3+/4+}$ states -as expected for the double exchange driven ferromagnetic metallic state- and no traces of anomalous manganese Mn$^{4+}$ or Mn$^{3+}$ ionic states that would have been a sign of charge localization in the material [25]. Similar experiments on (001)LCMO films invariably revealed evidence of electronic phase separation into dominating ferromagnetic metallic phase (Mn$^{3+/4+}$) and ferromagnetic insulator with distinguishable localized Mn$^{4+}$ and Mn$^{3+}$ states [13]. Therefore, these experiments suggested that the (110) films are electronically more homogeneous.

We will focus on the relationship between structural parameters and magnetic properties of (001) and (110)LCMO films with thicknesses (*t*) in the 8-150 nm range. Detailed structural analysis revealed that the (001) films are epitaxially clamped to the substrate and, within the explored *t*-range, the in-plane unit cell parameters do not change appreciably. In contrast, the in-plane interplanar distances in the (110) films gradually relax (contracting) at different rate along the in-plane [1-10] and [001] directions. Therefore, the resulting biaxial strain of (110) films is found to be much smaller than that of (001) films. We will show that the (110) films display, at a given thickness, higher T$_C$ and larger M$_S$ than their (001) counterparts. When analyzing the T$_C$ dependence of all films on biaxial strain it will be clear that whereas for relatively small biaxial strain ($\varepsilon_{bi}$ < 1.2 10$^{-2}$) –as observed in all (110) films- the elastic model of Millis *et al.* [6] allows describing the magnetic data, this is not the case for more strained films -typically (001) films-, so other effects must be invoked to account for these experimental observations. Moreover, the reduced $M_S$ found in (001) films, much more pronounced than in the (110) films, suggests the presence of more abundant non-ferromagnetic regions in (001) films. $^{55}$Mn-nuclear magnetic resonance has confirmed the presence of electronic phase separation in (001) films and its absence in (110) ones.

We argue that the observed variation of the magnetic properties of epitaxial manganite thin films versus thickness results from the combined effect of the elastic deformation of the lattice, which narrows the $e_g$ bandwidth and reduces the hopping probability that appears at relatively weak strain, and, on the other hand, the orbital ordering and subsequent phase separation occurring at larger strain. The broader conduction band in (110) films reduces the trend to phase separation. Implications for using (110) manganite electrodes in spintronic components are discussed.

## II. EXPERIMENTAL

Prior to the growth of manganites films, the (001) and (110)STO substrates were cleaned in ultrasonic baths and thermally treated (1000ºC for 2 h in air). Manganite films were simultaneously deposited onto (001) and (110)STO substrates by rf magnetron sputtering using a $La_{2/3}Ca_{1/3}MnO_3$ stoichiometric target; the substrate-holder was heated to 800ºC and the total chamber pressure was fixed to 330 mTorr (80% Ar, 20% $O_2$). At the end of the deposition, the chamber atmosphere was changed to 350 Torr of pure oxygen and the samples were annealed (at 800ºC) for 1 h before cooling down to room temperature. Samples were grown at a rate of 0.2 nm/min, thicknesses being from 8 nm to 150 nm were deduced from X-ray reflectometry measurements. X-ray diffraction (XRD) and reciprocal space maps (RSM's) were performed to determine the unit cell parameters. Some samples, prepared in cross section geometry (XTEM), were studied by high resolution TEM (HRTEM) in a Jeol J2010F microscope with a field emission gun operating at 200 keV.

X-ray photoemission spectra were recorded with an ESCALAB 210 spectrometer fitted with a hemispherical energy analyzer working at a constant pass-energy of 30 eV. The base pressure of the analysis chamber was approximately $10^{-10}$ Torr. Non-monochromatic Al-K$_\alpha$ radiation was used as the excitation source. Survey spectra and the following core levels were recorded: La3d$_{5/2}$, Ca2p, Mn2p, O1s and C1s. Binding energy (BE) calibration of the spectra was done by referencing the recorded peaks to the La3d$_{5/2}$ at 834.9 eV. Intensity ratios for the different photoemission signals have been calculated by estimating the area of the elastic photoemission peaks and by referring the obtained values to their relative sensitivity factors [26]. Sputtering with $Ar^+$ ions (2 keV, 60º grazing incidence, 3.5 minutes) was used to remove surface contamination.

Magnetic properties were studied using a commercial superconducting quantum interference device (SQUID, Quantum Design). Magnetic measurements were carried out with the magnetic field applied in-plane, for (001) samples being parallel to [100] direction and for (110) ones parallel to [001] direction. Temperature dependences of magnetization for all films were recorded at 5 kOe and hysteresis loops were measured at 10 K with a maximum magnetic field of 30 kOe.

Spin Echo NMR experiments were performed at 4.2 K, in zero external magnetic field, and at several values of the excitation rf field in order to optimize the signal intensity at each frequency. NMR spectra were obtained in the frequency range 300-450 MHz by plotting every 1 MHz an optimum spin echo signal intensity corrected for the intrinsic enhancement factor η and the frequency dependent sensitivity factor $\omega^2$, as described in details in Refs. 27 and 28.

### III. RESULTS

#### A. Structural data

In Fig. 2 we show typical θ-2θ XRD scans of films of nominal thickness $t = 85$ nm on (001) and (110)STO substrates. Data in Fig. 2 show that both films are fully textured with out-of-plane directions [001] and [110] respectively. No traces of other phases or textures are observed. We notice that the film reflections are shifted towards larger angles when compared to those of the substrate, thus indicating smaller out-of-plane lattice distances ($d_{001}$ and $d_{110}$, respectively) than those of the substrate. This results from the tensile strain imposed by the substrate on the in-plane interplanar distances of LCMO, as the STO unit cell is larger than that of the LCMO with a mismatch of $f = +1.1$ %. RSM's around asymmetric reflections were collected to determine in-plane and out-of-plane interplanar distances.

Fig. 3 shows the RSM collected around (103) reflection for two (001) films. The inspection of these plots immediately reveals that the in-plane interplanar distance ($d_{100}$) of the $t = 8$ nm LCMO film (Fig. 3a) is clamped to the substrate, and it remains practically fully strained up to $t = 80$ nm (Fig. 3b). Thus, the (001) films remain fully strained even for thickest (150 nm) one. Besides, the out-of-plane interplanar distance ($d_{001}$) of these films displays a weak, non-monotonic, variation when thickness increases.

In the (110)LCMO films, two unequivalent in-plane orthogonal directions exist: [001] and [1-10]. To obtain the interplanar distances along these directions ($d_{001}$ and $d_{1-10}$, respectively) RSM's around (222) and (130) reflections were collected. Corresponding data for two (110) films are shown in Fig. 4. Inspection of the data for the thinner (8 nm) film reveals that whereas along the [001] direction the film remains fully strained (a), this is not the case for the [1-10] direction (c) where a clear relaxation is observed. For the 80 nm film, in-plane relaxation is observed along both directions [001] (b) and [1-10] (d). Therefore, whereas the in-plane interplanar distances of the (001) films remain clamped to those of the substrate and thus films remain fully strained, the (110) films display a gradual and anisotropic relaxation when thickness increases.

The thickness dependence of the interplanar distances evaluated from RSM's for the (110) and (001) films are summarized in Fig. 5a. We first comment the data for (001) films (solid symbols). Virtually no relaxation occurs for the in-plane interplanar distance $d_{100}$ (solid triangles), which remains clamped to the substrate. The out-of-plane $d_{001}$ distance (solid rhombi) displays a non-monotonic behavior: it contracts for films up to a critical thickness $t_c \approx 20$ nm and expands slightly for the $t > 40$ nm films, remaining well below the bulk value (indicated by the dashed line) even for $t \approx 150$ nm. The persistence of a strained state in LCMO (and other manganites) epitaxial thin films has been reported earlier [5,13].

We turn now to (110) films. Data in Fig. 5a (open symbols) reflect the gradual structural relaxation along both in-plane directions [001] and [1-10] as well as the out-of-plane [110] one. It is clear that the in-plane distances $d_{001}$ and $d_{1-10}$ relax much faster than the corresponding in-plane $d_{100}$ in the (001) films. At $t \approx 150$ nm the relaxation is almost complete. Moreover, the relaxation along the [001] and [1-10] directions is different: for $t < 20$ nm, the lattice contracts faster along [1-10] direction than along [001] one. The out-of-plane $d_{110}$ distance displays a non-monotonic behavior: it contracts up to $t_c \approx 20$ nm but it expands markedly for a further increase of film thickness, approaching the bulk value for $t \approx 150$ nm.

Therefore, the response of (001) and (110) films to identical lattice mismatch imposed by the substrate is quantitatively different; however, a common trend is identified, namely the evidence of a critical thickness $t_c$ that separates the well-defined regions in the thickness dependence of the out-of-plane interplanar distance. To emphasize the existence of these two different structural regions, we have linearly extrapolated the thickness dependence of the out-of-plane distance from $t > t_c$ towards $t \rightarrow 0$. The dashed areas in Fig. 5a indicate a range of thicknesses where the out-of-plane distances are substantially larger that could be expected from their monotonic variation observed above $t_c$.

The evolution of the unit cell volume with thickness is depicted in Fig. 5b. The inspection of these data indicates several important characteristics: (i) the unit cell volume of the (001) and (110)LCMO films ($V_{001}$ and $V_{110}$ respectively) are substantially larger than the bulk value $V_{LCMO}$ (indicated by a dashed line in Fig. 5b); (ii) upon increasing thickness, $V_{001}$ (solid symbols) changes slightly remaining above the bulk value even for the thickest film; (iii) the gradual reduction of the unit cell volume is clearly pronounced for the (110) films: for $t > 30$ nm, $V_{110} < V_{001}$ and at $t \approx 150$ nm, $V_{110}$ reaches the bulk value. Data in Fig. 5b clearly illustrate that the assumption of constant unit cell volume when describing strain effects is not supported by the experimental data. Moreover, they also reveal that for film with thickness below $t_c$, the unit cell volume displays a rapid expansion which is more pronounced than what could be expected from a simple linear extrapolation towards $t \rightarrow 0$. Dashed areas in Fig. 5b illustrate these anomalous unit cell volume regions for (001) and (110) films, mimicking the corresponding regions in the out-of-plane interplanar distance evolution shown in Fig. 5a.

Samples grown on (001)STO substrate present an epitaxial relationship LCMO(001)[100]//STO(001)[100], as deduced from X-ray diffraction and HRTEM. In Fig. 6a we show a high resolution XTEM image of a 8 nm LCMO film along the [100] zone axis. It can be appreciated that, in the examined area, the film-substrate interface is abrupt with no evidence of misfit dislocations. The lattice parameters extracted from XTEM are in good agreement with those extracted from XRD data and confirm that (001)LCMO film grows coherently on (001)STO substrate. Similar experiments have been done on LCMO films grown on (110)STO. As expected, the epitaxial relationship is LCMO(110)[001]//STO(110)[001]. The high resolution images (Fig. 6b) do not show other relevant differences with respect to the sample grown on (001)STO than a greater roughness at the interface. No misfit dislocations have been observed in the considered region either. Rougher (110) interfaces can be expected on the basis of the polar nature of the (110)STO surface and its inherent tendency to reconstruction.

**B. Magnetic data**

The temperature dependent magnetization M(T) and the magnetization loops M(H) of t = 43 and 85 nm (001) and (110)LCMO films are shown in Fig. 7. The M(T) curves in Fig. 7a display a decrease of the magnetization and a lowering of the temperature onset ($T_C$) of ferromagnetic behavior as the thickness decreases. This trend has been reported earlier for (001) films of LCMO [3,13] and other $R_{1-x}A_xMnO_3$ manganites [1,2] grown on different substrates. The same trend is displayed by (110)LCMO films (Fig. 7c). However, there are two fundamental differences: the decay of $T_C$ when reducing thickness is much less pronounced and the ferromagnetic transition is sharper for (110) films than for (001) ones. These observations indicate that (110) films indeed present improved magnetic properties. This trend is also inferred and confirmed from the analysis of the hysteresis loops shown in Fig. 7b and d. Comparison of data for (001) and (110) films immediately evidences that the saturation magnetization ($M_S$) of the (001) films is smaller than that of the (110) films at each thickness. Data in Fig. 8, where the thickness dependence of $T_C$ and $M_S$ for all (001) and (110) films is summarized, clearly show that the (110)LCMO films have higher Curie temperature *and* saturation magnetization than (001)LCMO films of similar thickness.

To get an insight into the origin of this dissimilar behavior, we have performed $^{55}$Mn NMR experiments at T = 4.2 K. Fig. 9 shows two examples of the collected spectra corresponding to (001) (solid symbols) and (110) (open symbols) films of t = 85 nm (a) and 43 nm (b). We first note that in all the spectra the dominant line (around 375-380 MHz) comes from the mixed-valent $Mn^{3+/4+}$ state and its frequency corresponds $^{55}$Mn hyperfine field averaged due to double exchange driven fast motion of electron-holes, over manganese sites $Mn^{4+}$- $Mn^{3+}$. Consequently, the frequency of this line is determined by the ratio of $Mn^{3+}$ to $Mn^{4+}$ and varies proportionally with charge carrier density in the conducting band [30]. The resonant frequency increases with increasing film thickness and depends on the film orientation: it is larger for (110)LCMO than for (001) films at each thickness (ν(001) = 376 MHz and ν(110) = 380 MHz, for 85 nm, for instance). An increase of the resonant frequency with thickness is in agreement with the trend reported in Ref. 13. We note also that the resonant frequency is larger for the (110) film, which has a higher Curie temperature.

On the other hand, the spectra of the (001) films (see insets in Fig. 9) reveal the presence of a broad peak of small intensity appearing at lower frequency (around 315 MHz) and featuring a high value of NMR enhancement factor characteristic for ferromagnetic phase. A similar feature had been previously observed in the spectra of (001)LCMO films and attributed to the presence of hole-trapped (localized) ferromagnetic $Mn^{4+}$ state [13]. Moreover, in the present experiment, a clear, albeit tiny, asymmetry is observed on the high frequency side of the $Mn^{3+/4+}$ line revealing the presence of additional Mn state giving rise to the NMR $Mn^{3+}$ state producing a NMR signal in this frequency range. The frequency of this resonance (~ 400-420 MHz) corresponds to a $Mn^{3+}$ state [30] and thus it identifies the presence of localized ferromagnetic $Mn^{3+}$ ions in the (001) film. A sizable amount of localized $Mn^{4+}$ and $Mn^{3+}$ ions in (001) LCMO films detected by $^{55}$Mn NMR evidences the presence of ferromagnetic insulating phase with localized charges and consequently it can be considered as a signature of electronic phase separation taking place in these films [13]. Charge localization implies weakening of the double exchange coupling, and thus a reduction of the Curie temperature is foreseen for (001) films. Remarkably, the NMR spectrum of the (110) films does not show these localized $Mn^{3+}$ or $Mn^{4+}$ states, and thus the phase separation can be considered to be absent or, at least, much less abundant; accordingly, the Curie temperature (and magnetization) should approach ideal, bulk-like values, which we have indeed observed.

Under the appropriate measurement conditions, the total area under the NMR spectra is determined by the fraction of magnetic ions in the sample [31]. The integrated area under the spectra for (001) and (110) samples of various thicknesses have been measured and these values are summarized in Fig. 10. As expected, there is a gradual increase of the NMR intensity at T = 4.2 K as a function of t for both series of samples. However, there is a striking difference: the NMR intensity is always larger for the (110) samples than for the (001) ones. This observation clearly indicates that the ferromagnetic fraction in the (110) samples is always larger than that of the (001) ones. We note that the missing intensity would be related to $Mn^{m+}$ ions either antiferromagnetically coupled or paramagnetic, and located in a non-ferromagnetic phase, characterized by a very low NMR enhancement, which therefore contributes with a negligible intensity to the NMR spectrum of strong ferromagnetic signals. Extrapolating to zero the experimental NMR intensity, it has been evaluated the thickness of this non-ferromagnetic layer ($t_c$), being for (001) films $t_c^{001} = 5.5$ nm, whereas for (110) ones it is virtually zero, $t_c^{110} = 0$ nm. Therefore, from NMR experiments we conclude that ferromagnetism in (110)LCMO films appears to be more magnetically robust than that in (001)LCMO counterparts. This observation is in good agreement with the macroscopic magnetic properties reported above.

The observation of a ferromagnetic charge-localized signal in the NMR spectra of $La_{1-x}Ca_xMnO_3$ manganites has been viewed [32] as a signature of the presence of ferromagnetic insulating regions with A-type orbital ordering. As we will discuss in detail below, within this framework our results would imply that orbital ordering is reinforced in the most strained (001)LCMO films.

### C. X-ray photoelectron spectroscopy

We have investigated the chemical composition of the (001) and (110) films surface (after ion etching of around 0.5 nm) by X-ray photoelectron spectroscopy. In Figs. 11 (a and b) we show the La3d and Mn2p photoelectron emission spectra of 17 nm thick (001) and (110)LCMO films (solid and dashed lines, respectively). The recorded intensity has been normalized to the intensity of the La3d lines. From this figure, it can be immediately realized that the intensity of the Mn2p line is identical for (001) and (110) films, thus indicating the same Mn/La ratio for both samples. The spectra corresponding to the Ca2p lines are shown in Fig. 11c. Inspection of these spectra reveals that the intensity of the Ca2p spectrum of the (001) surface is larger than that corresponding to the (110) film, thus suggesting a relative Ca enrichment of the (001)LCMO surface when compared to (110)LCMO one.

Simon *et al.* [15] characterized by secondary ion mass spectrometry and energy electron loss spectroscopy (EELS) the chemical composition of strained LCMO films on STO(001) substrates, and they found Ca segregation to the surface. Similarly, the presence of a Sr-rich surface in strained (001)$La_{2/3}Sr_{1/3}MnO_3$ thin films, detected by XPS, was reported by Bertacco *et al.* [33]. Observation of Sr-rich layer at the surface of strained $La_{2/3}Sr_{1/3}MnO_3$ was documented by Maurice *et al.* [5] by using EELS. Our data show that the Ca enrichment at the surface of the (110) films is substantially less pronounced that for (001) films.

## IV. DISCUSSION

As previously mentioned, many studies were focused on thickness effects in manganite thin films grown on different substrates. The most common scenario attributes the variation of the Curie temperature to the elastic deformation of the perovskite lattice, subsequently changing the Mn-O-Mn bond-angle and bond-distances. However, evidences have been reported suggesting that strain effects are not enough to account for the experimental results. For instance, a depression of $T_C$ and $M_S$ has been observed in thin films grown on virtually matching substrates (NdGaO$_3$ and (Sr$_2$AlTaO$_6$)$_{1-x}$(LaAlO$_3$)$_x$) [7,13,34], and our accurate determination of the unit cell volume as a function of thickness clearly indicates that the assumption of elastic deformation of the unit cell under strain is not experimentally supported. On the other hand, the consideration of elastic effects as the unique driving force of the changes in manganite thin film magnetic properties neglects the common observation of their reduced saturation magnetization, which has been related to either a dead magnetic layer [2,13] or an electronic phase separation at the interface or extending deep into the film [13]. Therefore, the mechanism of suppression of $T_C$ and $M_S$ has not been yet definitively settled.

### A. Elastic effects

We have shown that the unit cell volume values $V_{001}$ and $V_{110}$ for LCMO films largely differ from the bulk value $V_{LCMO}$ (Fig. 5b) in our thickness range of study, and although they gradually approach $V_{LCMO}$ when increasing film thickness, for $t \leq 150$ nm, they always remain sensibly larger than $V_{LCMO}$. Thus, the unit cell volume is not preserved but changes with strain, indicating that a constant Poisson ratio can not be assumed to evaluate the biaxial strain ($\varepsilon_{bi}$). Instead, measurements of in-plane and out-of-plane interplanar distances are required for any thickness to determine $\varepsilon_{bi}$. Before presenting the dependence of $\varepsilon_{bi}$ on thickness, we would like to comment on the measured anisotropic in-plane relaxation of the interplanar distances in the (110)LCMO films. The LCMO lattice relaxes faster along the [1-10] direction than along the [001] direction for $t < 20$ nm (Fig. 5a). In fact, the measured strain ($\varepsilon$) and Young modulus (Y) determine the resulting stress ($\sigma = Y\varepsilon$). The radically different atomic sequence along [001] and [1-10] directions foresees the different Young moduli ($Y_{001} \neq Y_{1-10}$) and thus structural relaxation could be in-plane anisotropic. In Fig. 12a we plot the in-plane ($\varepsilon_{11}$ and $\varepsilon_{22}$) and out-of-plane ($\varepsilon_{33}$) strain coefficients of the (001) and (110) films, evaluated as the difference between interplanar distances for film and corresponding values for bulk samples (Fig. 5a). Fig.

12b shows the thickness dependences of the bulk ($\varepsilon_B$) and biaxial strains ($\varepsilon_{bi}$), calculated using $\varepsilon_B = 1/3\ (\varepsilon_{11} + \varepsilon_{22} + \varepsilon_{33})$, and $\varepsilon^{in}_{bi} = (1/2)\ (\varepsilon_{11} - \varepsilon_{22})$ and $\varepsilon^{out}_{bi} = (1/4)\ (2\varepsilon_{33} - \varepsilon_{11} - \varepsilon_{22})$.

To observe the influence of unit cell strain on $T_C$, we recall eq. (1). In this expression, there is a term related to unit cell compression given by $\alpha\varepsilon_B(t)$. For a film under a (thickness dependent) bulk strain $\varepsilon_B(t)$, the resulting stress is $\sigma_B \approx K\varepsilon_B(t)$, K being the bulk modulus elasticity [$K = (1/3)\ (c_{11} + 2c_{12})$]. The elastic constants of manganite films have been reported previously [35]. We have used the values of $c_{11}$ = 350 GPa and $c_{12}$ = 113 GPa for LCMO films determined as explained in Ref. 36. These elastic constants lead to K = 192 GPa, and the resulting bulk stress on thinner films ($\varepsilon_B \approx 0.5\%$) amounts $\sigma_B \approx 1$ GPa. The change of the unit cell volume associated to this compression should produce a change of $T_C$ comparable to what has been measured in experiments under hydrostatic pressure in LCMO crystals. From experiments performed on polycrystalline LCMO samples, $dT_C/dP \approx 0.9$ K/kbar [10]. Substituting this value and the corresponding bulk modulus K = 192 GPa we obtain $\alpha = (1/T_C)\ dT_C/d\varepsilon_B = K\ (1/T_C)\ dT_C/dP \approx 6.4$. From experimental data in Fig. 12b, the maximal bulk strain occurring in the thinner films ($\varepsilon_B = 0.5$ %) will induce a reduction of $T_C \approx 3$ %. The $T_C(t)$ dependence (Fig. 8) clearly shows stronger variation, thus indicating that bulk strain is not enough to account for the observed thickness dependence of $T_C$.

The biaxial strain effect on $T_C(t)$, is presented in Fig. 13, where $T_C$(150nm) is the Curie temperature of the 150 nm film and $\varepsilon^2_{bi}$ accounts for the total biaxial strain, $\varepsilon^2_{bi} = \varepsilon^{in\ 2}_{bi} + \varepsilon^{out\ 2}_{bi}$, measured for each film. Data appear concentrated around two strain regions, reflecting the fact that (110) films (circles) have much smaller biaxial strains than (001) ones (squares). Inspection of data in Fig. 13 signals that for (110) films $T_C / T_C$(150nm) is roughly linear on $\varepsilon^2_{bi}$ as predicted by eq. (1), thus suggesting that for (110) films elastic strain effects could be at the origin of the suppression of $T_C$ of strained films. However, the data for most strained (001) films dramatically differ from the linear behavior. In fact, data in Fig. 13 indicate that the Curie temperature of the (001) films changes without a significant variation of biaxial strain. In order to illustrate that the biaxial strain dependence of $T_C / T_C$(150nm) is not greatly affected by the presence of the bulk strain contribution, in Fig. 13 we also include data corrected by the bulk strain contribution ($T_C / T_C$(150nm) + $\alpha\varepsilon_B$) (open symbols). We thus conclude that the strong dependence of $T_C$ on thickness observed in the thinner ($t <$ 85 nm) (001) films is not dominated by the elastic strain deformation of the manganite lattice. However, one can notice that the linear dependence of the $T_C / T_C$(150nm) on ($\varepsilon^2_{bi}$) for (110) films can be extrapolated to higher strain region (solid line in Fig. 13) and overlaps with data recorded for the less strained (001) films. This observation indicates that two different mechanisms control the $T_C$ suppression: the elastic strain-induced deformation of the lattice and subsequent reduction of the hopping probability between 3d-$e_g$ orbitals, as predicted by the model of Millis *et al.* [5], dominate the $T_C(t)$ dependence of the films with biaxial strain $\varepsilon^2_{bi} < 1.6 \; 10^{-2}$, whereas a non-elastic mechanism, related to the occurrence of phase separation (see below), is more relevant for more strained films, typically the (001)LCMO ones.

### B. Phase separation

Key observations to evidence the major relevance of phase separation in the most strained (001) films are the results obtained from NMR experiments and saturation magnetization: (i) The NMR data show clear evidences of localized $Mn^{4+}(Mn^{3+})$ state correlated with the presence of phase separated ferromagnetic insulator in (001)LCMO films and a smaller fraction of ferromagnetic ions than in (110) films and (ii) the saturation magnetization of the (001) films is found to be smaller than that of (110) ones of similar thickness.

To get more an insight into the possible origins of the observed phase separation and their relation to strain, we discuss separately two different scenarios: a) Coexistence of orbital-ordered phases (and subsequent electronic charge localization) with orbital disordered ones, and b) Chemical segregation.

**a)** It has been reported that biaxial strain in manganite thin films can trigger different orbital ordering such as antiferromagnetic A-type or antiferromagnetic C-type depending on the sign (tensile or compressive) of the strain [18-20,37]. For instance: tensile in-plane strain on a (001) film stabilizes the $x^2-y^2$ orbital levels while the subsequent out-of-plane compression pushes up the $z^2$ orbital level, thus favoring electron occupancy of the $x^2-y^2$ orbitals and eventually leading to A-type orbital ordering [38]. In Fig. 14a, we sketch the relevant orbitals for a (001)LCMO film and the resulting energy splitting under a biaxial tensile strain. In Fig. 14b we depict the corresponding orbital arrangement for a (110) film similarly strained and the subsequent energy splitting of the $e_g$ orbitals. In the (110) case, the in-plane $z^2$ orbital level becomes stabilized by the tensile epitaxial strain and, more relevant, the $x^2-y^2$ orbitals are not significantly lifted in energy. Therefore, the energy splitting between $z^2$ and $x^2-y^2$ orbitals in this case should be much smaller than that occurring in the (001) films and thus the driving force towards antiferromagnetic orbital ordering should be reduced in (110) films. Consequently, the degenerated metallic and ferromagnetic orbital (F-type) ordering in (110) films should be more robust than in (001) films. This agrees with our magnetization and NMR observations. More explicitly, Fang *et al.* [38] have shown that the tetragonal distortion of the lattice is a measure of the strength of the F-type to A-type orbital ordering. To show this effect, we include in Fig. 15 the tetragonality ($\tau$) of the LCMO unit cell as a function of thickness, calculated from structural data (Fig. 5a) as $\tau^{001} = d_{001}/d_{100}$ for (001) films and $\tau^{110} = d_{001}/(d^2_{1-10} + d^2_{110})^{1/2}$ for (110) ones. Data in Fig. 15 indicate that the driving force towards the A-type ordering is stronger in the (001) (close symbols) than in the (110) films (open symbols), as clearly seen when comparing the film tetragonality values $\tau^{001}$ and $\tau^{110}$ to the expected ideal cubic F-type environment ($\tau = 1$, dashed line).

The coexistence of A-type and F-type phases, stronger in the (001)LCMO films, implies that the (001) ones can not be longer viewed as homogeneous but rather as a composite of A and F nanoclusters whose concentration may depend on and evolve with temperature. Indeed, the broader ferromagnetic transition observed for (001)LCMO films illustrates this phase separation.

So far, phase separation has been described and attributed to strain acting on the film. However, recently Brey [21] has shown that even in the absence of strain, at the interface between an optimally doped (001) epitaxial manganite thin film and an insulator (for example, $SrTiO_3$), there is a charge depletion that weakens the ferromagnetic double exchange coupling and thus the interface layers are unstable against antiferromagnetic correlations.

Terminating planes in (001)STO can be $TiO_2$ and SrO, both being electrically neutral. In contrast, the $(R_{1-x}A_x)O$ and $(MnO_2)$ layers of the upper growing manganite $R_{1-x}A_xMnO_3$ are charged: +(1-x) and –(1-x), respectively. Therefore, it is clear that due to the build-in potential at interfaces, charge redistribution within the manganite layer may occur, modifying their properties. In contrast, we note that the (110)STO free surfaces (SrTiO and $O_2$), are not neutral but charged: +4 and -4, respectively. However, in this case epitaxial stacking of a $R_{1-x}A_xMnO_3$ oxide on the (110)STO substrate would not lead to any polarity difference as the growing layers $(R_{1-x}A_xMnO$ and $O_2)$ are also polar (+4 and -4) for any doping level. We thus conclude that even in the absence of strain effects, one could also foresee that (110)LCMO films are less prone to display orbital ordering and charge separation effects.

We notice that both strain and electrostatic effects reinforce antiferromagnetic orbital ordering reducing the total ferromagnetic response, and thus discrimination between both effects is ambiguous. However, we do expect electrostatic effects to be confined to few unit cells close to the interface [21] rather than ~ 5 nm deep as indicated by the NMR data (see Fig. 10).

**b)** XPS experiments have revealed a relative surface enrichment by divalent alkaline $Ca^{2+}$ in (001) films respect to (110) ones. As previously mentioned, some reports on $Ca^{2+}$ substituted manganites [15,34] have shown that phase segregation can occur through the interchange of $Ca^{2+}$ and $La^{3+}$ sites, being found to cause a $Ca^{2+}$-enrichment of LCMO free surfaces [15]. We note that a decrease of Ca-contents in the bulk of $(La_{1-x}Ca_x)MnO_3$ film and/or eventually at the film/substrate interface leads to an expansion of the unit cell due to the larger ionic radius of $La^{3+}$ ($\approx$ 1.16 Å) compared to that of $Ca^{2+}$ ($\approx$ 1.12 Å). Therefore, we could expect that the elastic energy required to accommodate lattice mismatch (tensile in the present case) can be reduced by compositional changes. In the present case, enrichment of the La/Ca ratio deep into the film is foreseen. This process would lead to $Ca^{2+}$ enrichment at the film surface in agreement with our XPS measurements. A similar process would also be operative in $La_{1-}$ $_x$Sr$_x$MnO$_3$ thin films on STO substrates, which are also under tensile strain. In fact, as reported by Maurice *et al*. [5] and Bertacco *et al*. [33], the surface of La$_{2/3}$Sr$_{1/3}$MnO$_3$ films on (001)STO were found to be Sr-rich. Simon *et al.* [15] reported a systematic EELS study of the interfaces in epitaxial films of La$_{2/3}$Ca$_{1/3}$MnO$_3$ and La$_{2/3}$Ba$_{1/3}$MnO$_3$ on (001)STO substrates and found that the film-substrate interface is alkaline depleted in the case of films which are under tensile strain.

One could thus speculate that a driving force for chemical phase separation is the reduction of the elastic energy imposed by the heteroepitaxial growth. As, in average, the films have the appropriate hole doping (x = 1/3) for optimal ferromagnetic properties, the strain-induced compositional changes can only produce a depression of them, eventually creating non-ferromagnetic regions.

Before concluding, it is worth to comment on the remarkable observation that the unit cell parameters of thinner (001) and (110) films (dashed areas in Fig. 5) anomalously increase when film thickness reduces below $t \sim 20$ nm. In low thickness LCMO films, de Andrés *et al*. [39] reported different crystal symmetry. Y. L. Qin *et al.* [40] also showed that thinner LCMO films on STO had a tetragonal symmetry instead of the orthorhombic observed in thicker ones. Our HREM images do not allow us to confirm that the same change of symmetry occurs in our thinner films, although the fact that a similar expansion of the unit cell was observed by de Andrés *et al.* suggests that this symmetry change can be the cause. However, as the thinner (001) and (110) films have very different strain effects and T$_C$'s, the change of symmetry can not be directly related to the magnetic properties of the ferromagnetic phase. We do not exclude that other defects can also promote the chemical or electronic inhomogeneities. Indeed, it has been previously shown that some other more trivial effects, such as the atomic quality of the surfaces of the underlying well-matching substrates (Sr$_2$AlTaO$_6$)$_{1-x}$(LaAlO$_3$)$_x$ for (001)LCMO grown onto them, has a significant roll onto the strength of the substrate-induced phase separation [34]. However, the systematic variations of all structural and magnetic properties with thickness reported here in the LCMO/STO system indicate that those effects are not relevant in the present case.

## V. CONCLUSION

We have shown that (110)LCMO films grown on (110)STO substrates display systematically higher Curie temperature and magnetization compared to the (001)LCMO films grown on (001)STO substrates simultaneously. It turned out that, due to differences in their elastic constants, the (110)LCMO films relax earlier than the (001)LCMO ones and thus the (110) films display a progressively reduced biaxial strain. The analysis of the dependence of the ferromagnetic ordering temperature on the biaxial strain has revealed that the elastic deformation of the manganite lattice can describe the response of the less strained films – typically the (110) ones- whereas other effects should be invoked to account for the properties of the most strained films–typically (001) ones. We have shown that phase separation, likely triggered by orbital ordering, may account for both the suppression of $T_C$ and the higher concentration of non-ferromagnetic and non-metallic regions in the (001) films than in (110) films. Although strain and charge depletion effects associated to polarity discontinuities at interfaces both reinforce antiferromagnetic orbital ordering, we have argued that the former dominates. Other effects, including chemical segregation, namely Ca-enrichment of the free surface of most strained films, can also be understood as a possible mechanism of releasing elastic energy at the cost of depressing their magnetic properties. In any event, it is clear that (110) films and (110) interfaces of manganite epitaxial films are magnetically more robust and thus they can be used to fabricate magnetic tunnel junctions with improved properties.

**ACKNOWLEDGEMENTS**

Financial support by the MEC of the Spanish Government (projects NAN2004-9094-C03 and MAT2005-5656-C04), by the European Community (project MaCoMuFi (FP6-03321) and FEDER), by the Ministry of Sciences and High Education of the Polish Government (Project No. 1 P03B 123 30), and the CSIC (Project 2006PL0021) is acknowledged. We are very thankful to L. Brey for illuminating discussions.

*Electronic address: icanerin@icmab.es

**Figures**

**Figure 1**

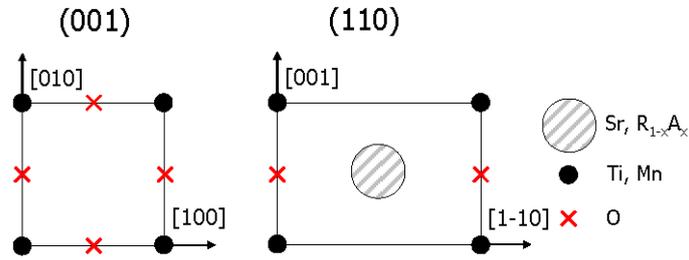

**Fig. 1** Sketch of the atomic arrangement in SrTiO$_3$ and (R$_{1-x}$A$_x$)MnO$_3$ unit cells for: left, (001)TiO$_2$ and MnO$_2$ planes, and right, (110)SrTiO$^{4+}$ and (R$_{1-x}$A$_x$)MnO$^{4+}$ planes, correspondingly.

**Figure 2**

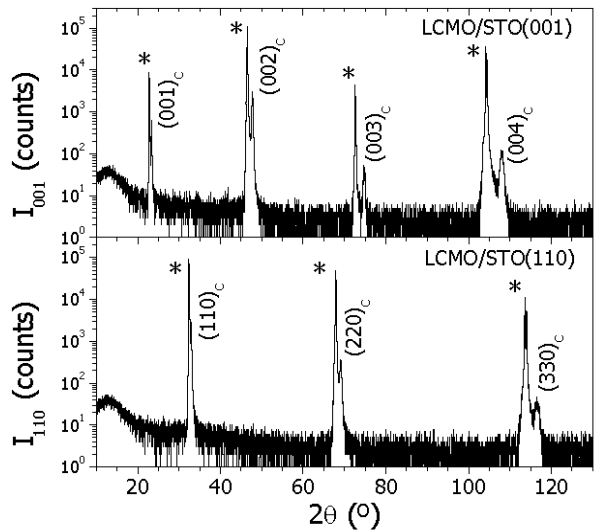

**Fig. 2** θ-2θ XRD scans of the nominal $t$ = 85 nm (001) (top panel) and (110) (bottom panel) films. The asterisks mark the substrate peaks.

**Figure 3**

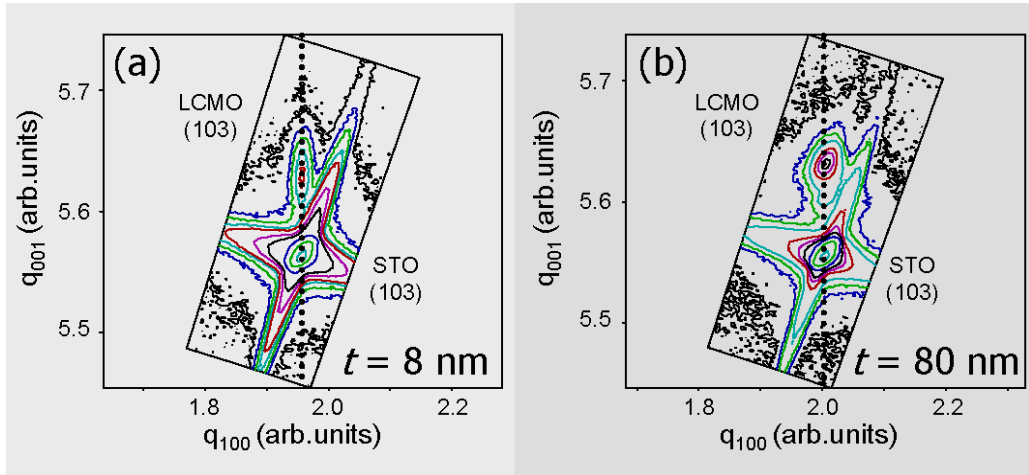

**Fig. 3** Reciprocal space maps collected around the (103) reflections for (001)LCMO films of $t$ = 8 nm (a) and 80 nm (b).

**Figure 4**

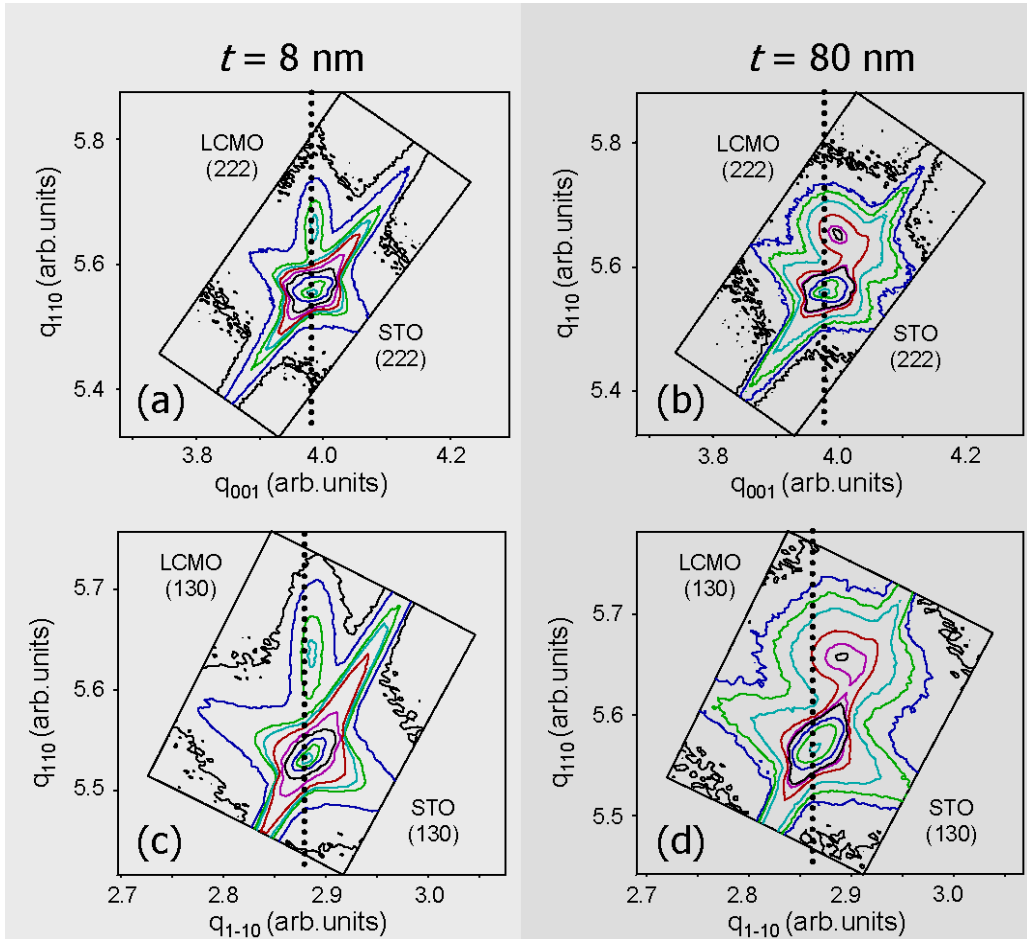

**Fig. 4** Reciprocal space maps around (222) (a, b) and (130) (c, d) reflections for (110)LCMO films. Data in (a) and (c) correspond to thinner ($t$ = 8 nm) film, in (b) and (d) to thicker ($t$ = 80 nm) film.

**Figure 5**

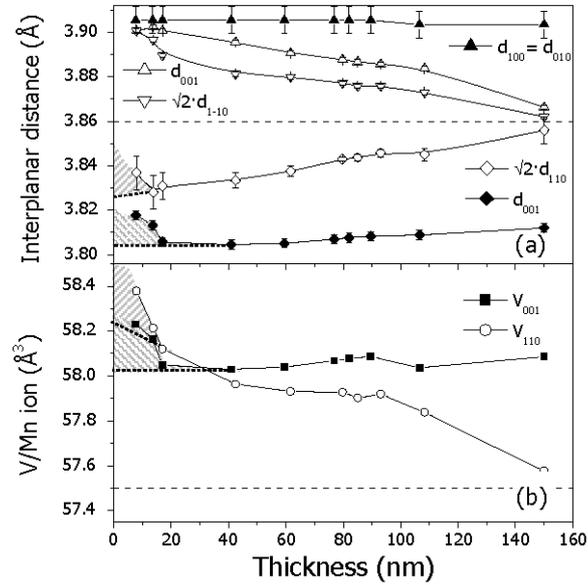

**Fig. 5** Thickness dependence of (a) interplanar distances and (b) unit cell volume for the (001)LCMO and (110)LCMO films. The dashed lines indicate the pseudocubic unit cell parameter (a) and unit cell volume (b) of bulk LCMO.

**Figure 6**

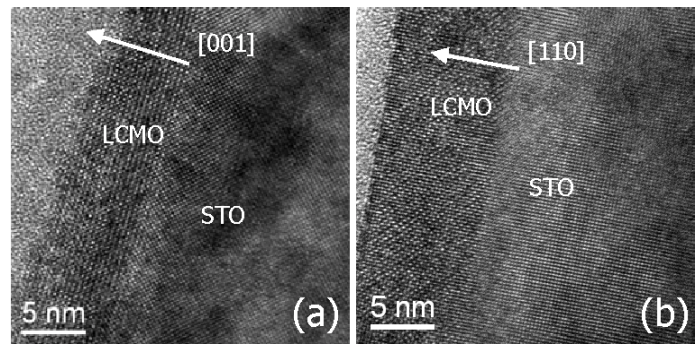

**Fig. 6** High resolution images of the interface of LCMO samples grown on: (a) (001)STO substrate along the [100] zone axis and (b) (110)STO substrate along the [001] zone axis.

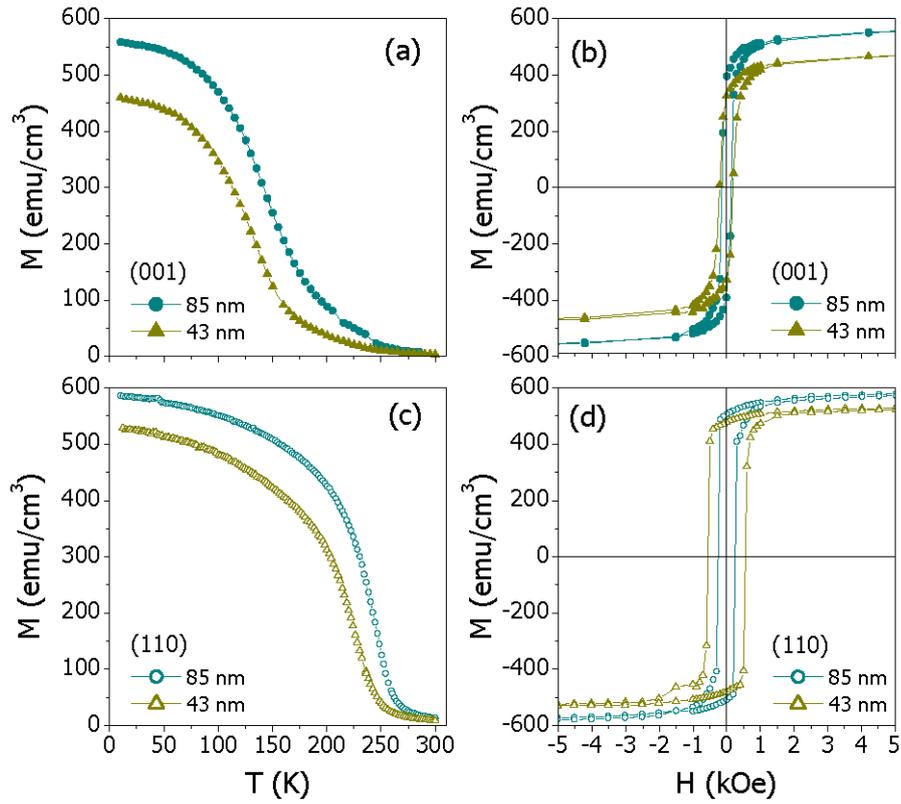

**Fig. 7** Magnetization *vs* temperature (measured at 5 kOe) of (001) (a) and (110) (c) films of thickness 43 nm (triangles) and 85 nm (circles). Field was applied in-plane along [100] direction for the (001) films and along [001] one for the (110) films. Magnetization *vs* magnetic field loops (b and d) measured at 10 K for the same films as in (a and c), respectively.

**Figure 8**

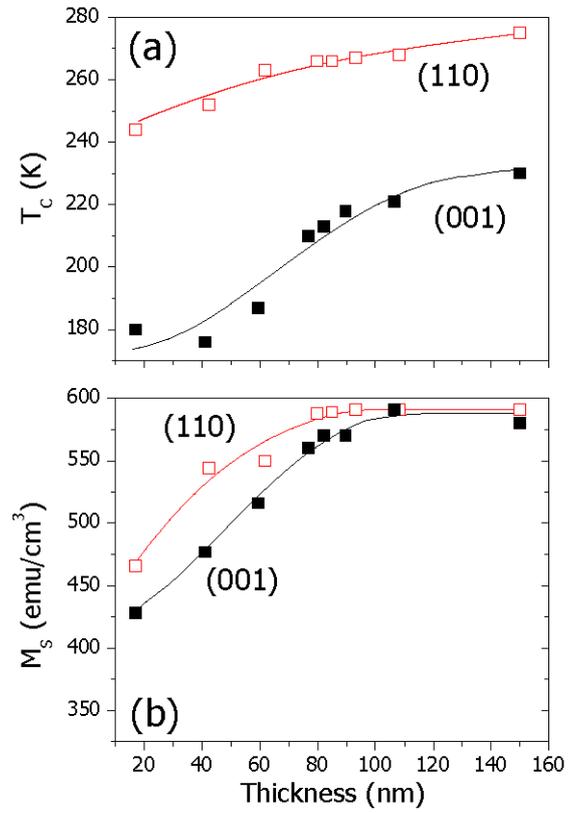

**Fig. 8** Curie temperature $T_C$ (a) and saturation magnetization $M_S$ (b) of (001) and (110) LCMO films (closed and open symbols, respectively) as a function of film thickness. Lines are guides for the eye.

**Figure 9**

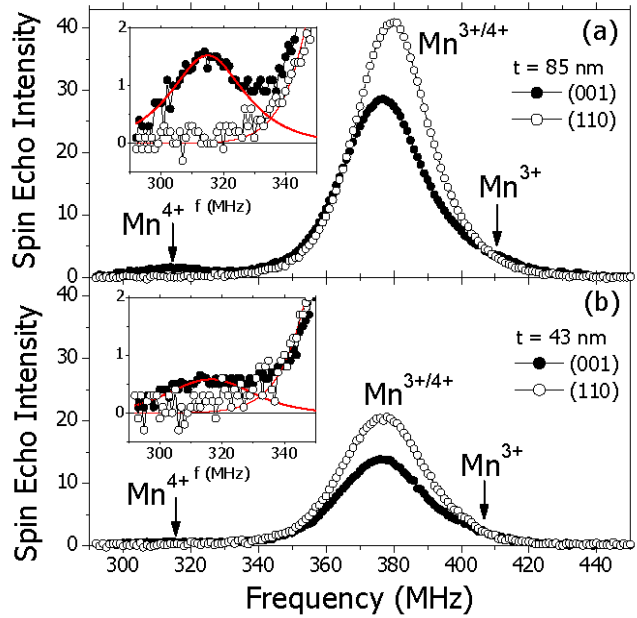

**Fig. 9** $^{55}$Mn NMR spectra of the (001) (solid symbols) and (110) (open symbols) films of thickness $t$ = 85 nm (a) and 43 nm (b) at T = 4.2 K. Insets show a zoom for low frequency data.

**Figure 10**

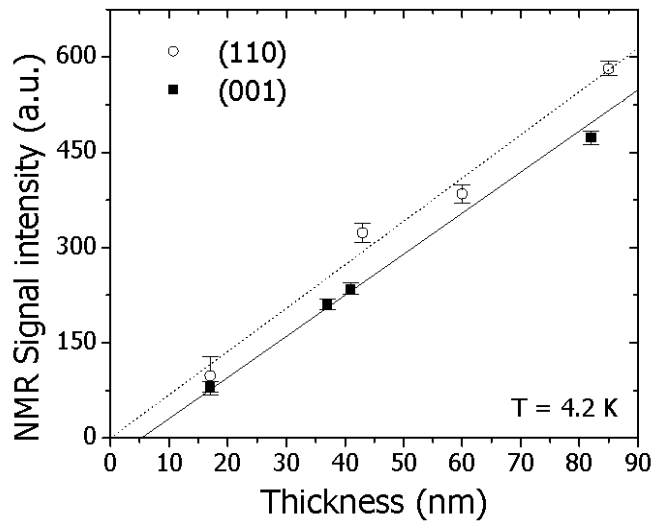

**Fig. 10** Thickness dependence of the integrated intensity of the NMR spectra for (001) (solid symbols) and (110) (open symbols) LCMO films at T = 4.2 K.

**Figure 11**

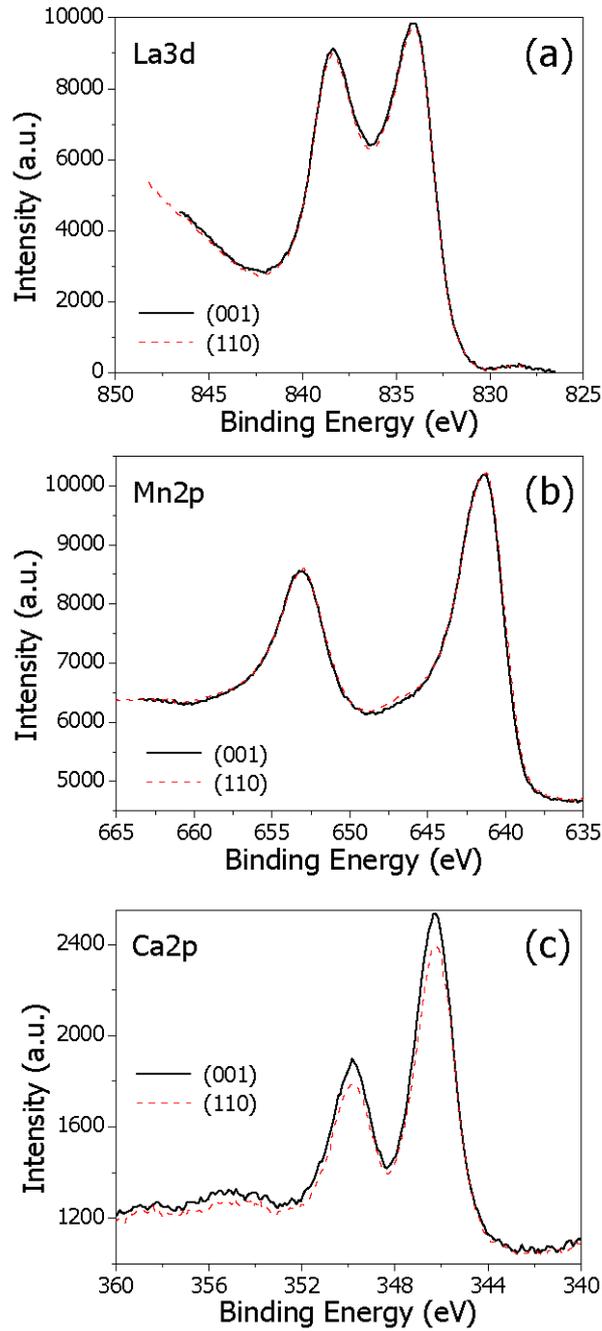

**Fig. 11** La3d (a), Mn2p (b) and Ca2p (c) photoelectron emission spectra for 17 nm (001) and (110)LCMO films (solid and dashed lines, respectively). The spectra were recorded after a soft ion etching. Normalization was performed using the La3d line.

**Figure 12**

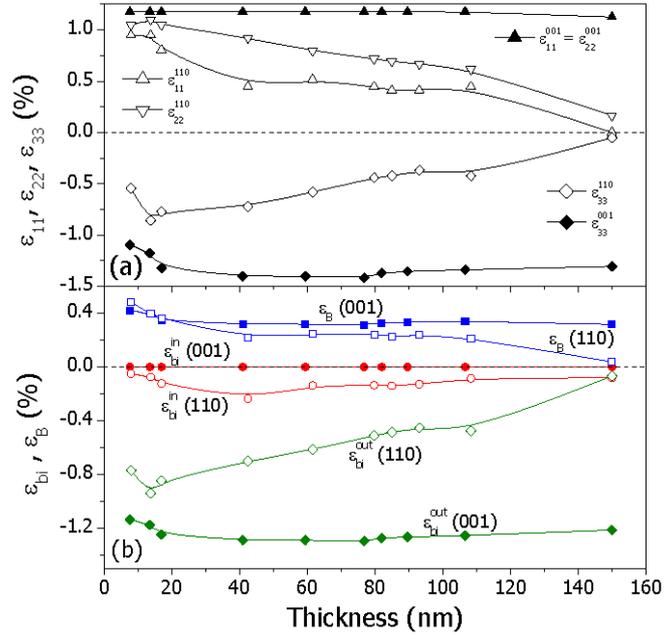

**Fig. 12** (a) Evaluated strain $\varepsilon_{11}$, $\varepsilon_{22}$ and $\varepsilon_{33}$ as a function of thickness for the (001) and (110)LCMO films. They correspond for (001) (solid symbols) to $\varepsilon_{11} = \varepsilon_{100} = \varepsilon_{010} = \varepsilon_{22}$ and $\varepsilon_{33} = \varepsilon_{001}$ whereas for (110) films (open symbols) to $\varepsilon_{11} = \varepsilon_{001}$, $\varepsilon_{22} = \varepsilon_{1\text{-}10}$ and $\varepsilon_{33} = \varepsilon_{110}$. (b) Bulk strain $\varepsilon_B$ (squares) and biaxial strain in-plane (circles) and out-of-plane (rhombi) for (001) (solid symbols) and (110) (open symbols) films as a function of film thickness.

**Figure 13**

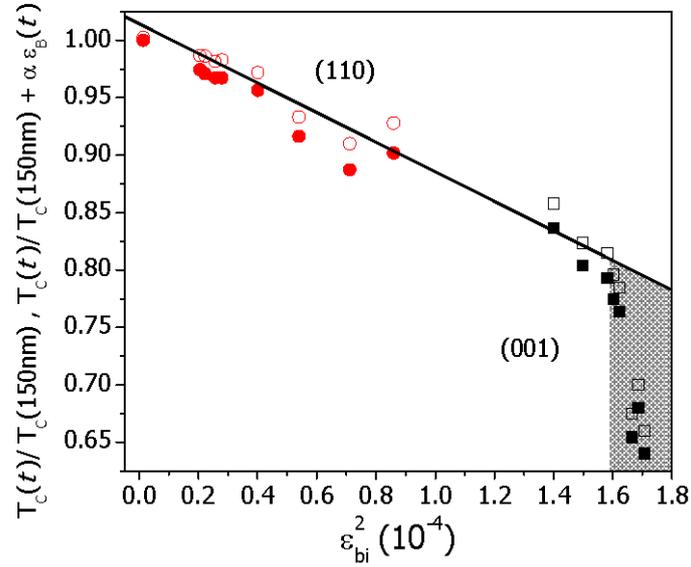

**Fig. 13** Dependence of the Curie temperature $T_C / T_C(150nm)$ (closed symbols) on the square of the biaxial strain $\varepsilon^2_{bi}$. Open symbols correspond to the $T_C / T_C(150nm)$ after correction for the bulk strain contribution (see text).

**Figure 14**

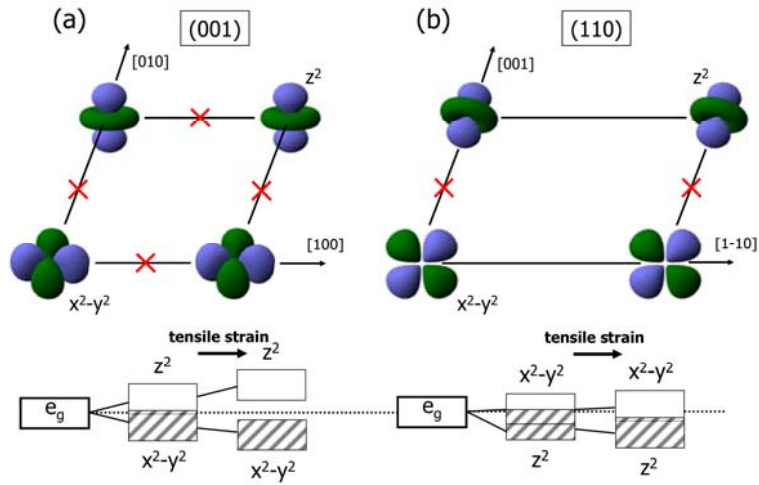

**Fig. 14** Sketch of the orbital arrangement (top panel) and energy levels (bottom panel) of $e_g$ orbitals in tensily strained (001) (a) and (110) films (b). For simplicity, at each Mn site it is shown only one of the $e_g$ orbitals and oxygen sites have been represented by crosses.

**Figure 15**

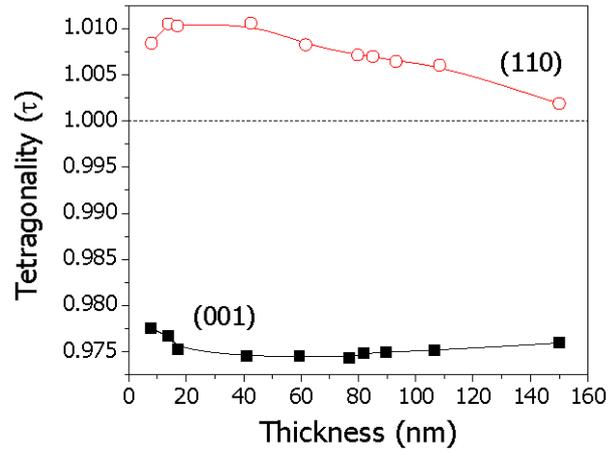

**Fig. 15** Tetragonality ($\tau$) of (001) (solid symbols) and (110) (open symbols) unit cell of LCMO films as a function of film thickness (see text).